\documentclass{elsart}

 \usepackage{epsfig}

\usepackage{amssymb}

\begin{document}

\begin{frontmatter}



\title{Hadron Correlations and Parton Recombination}


\author{R.~J.~Fries}

\address{School of Physics and Astronomy, University of Minnesota \\
Minneapolis, MN 55455\thanksref{fn1}}
\ead{rjfries@comp.tamu.edu}

\thanks[fn1]{Current affiliations: Cyclotron Institute and Department 
of Physics, Texas A\&M University, College Station TX 77801; \\
RIKEN/BNL Research Center, Brookhaven National Laboratory, Upton NY 11793}

\begin{abstract}
Parton recombination has been found to be an extremely useful model to
understand hadron production at the Relativistic Heavy Ion Collider. 
It is particularly important to explore its connections with hard processes.
This article reviews some of the aspects of the quark recombination model 
and places particular emphasis on hadron correlations.
\end{abstract}

\end{frontmatter}

\section{Introduction}

External probes are valuable tools for testing the properties of plasmas. 
In relativistic heavy ion physics hard QCD processes and electromagnetic 
processes take this role and are are supposed to unravel the secrets of 
the quark gluon plasma (QGP). Indeed the first
round of experiments at the Relativistic Heavy Ion Collider (RHIC) found
impressive evidence for jet quenching, a dramatic energy loss of the leading 
jet particle \cite{PHENIX:03pi0,star:06id}. Our interpretation of such 
results relies on the assumptions we make about the applicability of 
perturbative quantum Chromodynamics (pQCD). The view that momentum transfers 
of 1 to 2 GeV/$c$ are sufficient is supported by the success of 
next-to-leading order pQCD calculations for $p+p$ collisions at RHIC 
energies. However the answer does depend heavily on the process under
consideration.

Indeed we have now firm experimental evidence that in Au+Au collisions at RHIC
most hadrons are not produced from the usual jet fragmentation mechanism up 
to rather high transverse momenta $P_T \approx 6$ GeV/$c$. This makes
perturbative hadron production an even rarer process than most of us
expected. We want to call hadron production perturbative if it originates
from fragmentation of a jet.
Of course jets interact heavily with the surrounding medium. In the past it was
often assumed that the quenching leads to a reshuffling 
of energy within the jet cone or a transfer of energy from the jet into 
the medium without completely destroying the jet. As it turns out, the
number of hadrons from surviving jets at intermediate $P_T$ is drowned out
by the number of hadrons hadronizing from the bulk fireball via recombination
of quarks \cite{FMNB:03prl,FMNB:03prc,GreKoLe:03prl,HwaYa:02}.

The hints from experimental data can be divided into three categories. First,
one-particle inclusive hadron spectra show that the chemical composition 
of the hadrons produced between a $P_T$ of 2 and 6 GeV/$c$ is not compatible
with jet fragmentation. Most notable is the enhancement of baryons as seen
in the missing suppression of the nuclear modification factors $R_{AA}$ 
for baryons and in the various anomalously large baryon/meson ratios 
\cite{PHENIX:03ppi,STAR:03llbar,STAR:03v2}. These results already hint toward
a chemically equilibrated source of these hadrons.

Second, the elliptic flow coefficient $v_2$ follows to very good accuracy
a scaling law with respect to the number of valence quarks 
\cite{STAR:03v2,PHENIX:03v2}. This is very direct evidence that the 
observable $v_2$ is shaped in a deconfined quark phase.
A third emerging group of data involves dihadron correlations or associated
hadron yields. This data is much more difficult to describe (in any model)
than the single hadron spectra. Although jet-like correlations are present
at intermediate transverse momenta, the shape of these correlations and
the chemical composition can differ strongly from the vacuum 
\cite{STARcorr:02,phenix:04corr}.

The $\phi$ meson has long been discussed as a good test for the 
validity of the recombination model \cite{NMABF:03}. $\phi$ mesons are as 
heavy as protons but data from RHIC now impressively confirm that the 
elliptic flow and nuclear modification factor are very similar to those 
for kaons \cite{PalCai:05phi}. We can now be very sure that the meson vs 
baryon signature at intermediate $P_T$ is very robust and can neither 
be explained by an extrapolation of hydrodynamics nor by perturbative 
jet production and fragmentation.

\section{Recombination}

In central heavy ion collisions a hot and dense fireball of deconfined quarks 
and gluons is created.
In the recombination model one postulates the existence of thermalized parton 
degrees of freedom at the phase transition temperature $T_c$ which 
recombine or coalesce into hadrons. It has been found to be sufficient 
to consider the lowest Fock state in each hadron, the valence quarks, 
which are given constituent masses around 300 MeV.

The spectrum of hadrons can be calculated starting
from a convolution of Wigner functions \cite{FMNB:03prc}. 
For a meson with valence (anti)quarks $a$ and $b$ we have
\begin{eqnarray}
  \label{eq:reco}
  \frac{d^3 N_M}{d^3 P} &=& \sum_{a,b} \int\frac{d^3 R}{(2\pi)^3} 
  \int\frac{d^3 qd^3 r}{(2\pi)^3} \nonumber \\
  && \times W_{ab}\left(\mathbf{R}-
  \frac{\mathbf{r}}{2},\frac{\mathbf{P}}{2}-\mathbf{q}; \mathbf{R}+
  \frac{\mathbf{r}}{2},\frac{\mathbf{P}}{2}+\mathbf{q} \right)
  \Phi_M (\mathbf{r},\mathbf{q}).
\end{eqnarray}
Here $W_{ab}$ is the 2-particle Wigner function for partons $a$, $b$ and 
$\Phi_M$ is the Wigner function of the meson. The sum runs over all
possible parton quantum numbers. For simplicity
the parton Wigner function is usually approximated by a product
of single particle phase space distributions $W_{ab}=w_a w_b$.
Several slightly different implementations of this formalism have been 
discussed in the literature \cite{FMNB:03prc,GreKoLe:03prl,HwaYa:02}.
See \cite{Fries:04qm} for earlier reviews.

\begin{figure}
\begin{center}
\epsfig{file=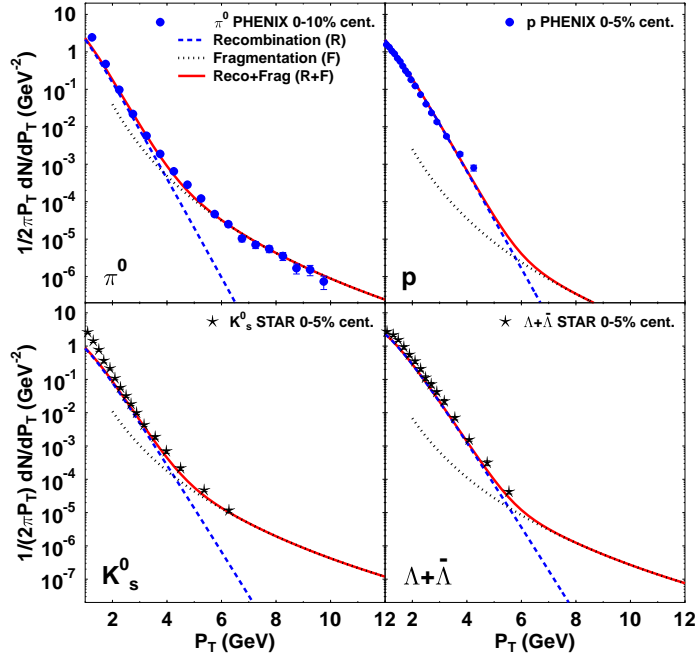,width=10cm}
\caption{\label{fig:spectra} Spectra of $\pi^0$, $p$, $K_0^s$ and $\Lambda+\bar
  \Lambda$ as a function of $P_T$ at midrapidity in central Au+Au collisions
  at $\sqrt{s}=200$ GeV \cite{FMNB:03prc}. Dashed lines are hadrons from
  recombination of the thermal phase, dotted line is pQCD with energy loss,
  solid line is the sum of both contributions. Data are from PHENIX ($\pi^0$, 
  $p$) \cite{PHENIX:03pi0,PHENIX:03ppi} and STAR ($K_0^s$, 
  $\Lambda+\bar\Lambda$) \cite{STAR:03llbar}.}
\end{center}
\end{figure}

A very good description of hadron spectra and hadron ratios measured at
RHIC can be achieved by combining hadron production from recombination 
for intermediate transverse momentum with a perturbative calculation using
fragmentation and energy loss in the medium \cite{FMNB:03prl,FMNB:03prc}.
Fig.\ \ref{fig:spectra} shows the $P_T$ spectrum of $\pi^0$, $p$, $K_0^s$ and 
$\Lambda+\bar\Lambda$ in central Au+Au collisions obtained in 
\cite{FMNB:03prc}. The agreement with data is very good 
for $P_T > 2$ GeV/$c$. We note that the hadron spectra exhibit an 
exponential shape up to about 4 GeV/$c$ for mesons and up to about 
6 GeV/$c$ for baryons, where recombination of thermal quarks dominates. 
Above, the spectra follow a power-law and production is dominated by 
fragmentation.

\begin{figure}
\begin{center}
\epsfig{file=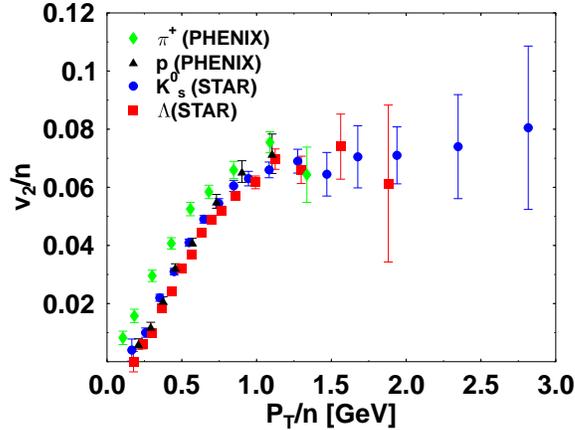,width=8cm}
\caption{\label{fig:v2} Elliptic flow $v_2$ for $\pi^+$, $p$, $K_0^s$ and
$\Lambda$ as a function of $P_T$ scaled by the number of valence quarks $n$ 
vs $P_T/n$. The data follows a universal curve, impressively confirming the 
quark scaling law predicted be recombination. Deviations for the pions are 
discussed in the text. Data are taken from PHENIX 
($\pi^+$, $p$) \cite{PHENIX:03v2} and STAR ($K_0^s$, $\Lambda$) 
\cite{STAR:03v2}.}
\end{center}
\end{figure}

Let us now assume the parton phase exhibits elliptic flow 
$v_2^{\mathrm p}(p_T)$. Recombination makes a prediction for elliptic flow 
of any hadron species after recombination which is precisely the scaling
law mentioned  before \cite{Voloshin:02,FMNB:03prc} 
\begin{equation}
  \label{eq:v2}
  v_2(P_T) = n v_2^{\mathrm p}(P_T/n)
\end{equation}
involving the number $n$ of valence quarks of the hadron. 
Fig.\ \ref{fig:v2} shows the measured elliptic flow $v_2$ for several
hadron species in a plot with scaled axes $v_2/n$ vs $P_T/n$. All data
points (with exception of the pions) fall on one universal curve.
The effect that pions are shifted to lower $P_T$ has been understood 
\cite{GreKo:04rho}. The quark scaling law is an impressive confirmation 
of the recombination model. It shows that the relevant degrees of freedom 
at early times in the collision are partons and it proves that these
partons behave collectively.

\begin{figure}
\begin{center}
\epsfig{file=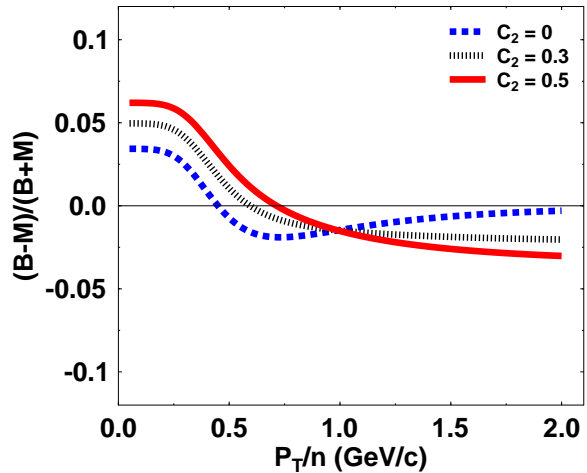,width=8cm}
\caption{\label{fig:hf} Scaling violation variable $A=(B-M)/(B+M)$ for
three different probabilities $C_2$ to have an additional parton beyond
the valence quark structure. Wave functions of finite width have been
used which lead to a scaling violation even in the case $C_2 = 0$ of 
pure valence quark recombination \cite{MFB:05}.}
\end{center}
\end{figure}

It is an interesting question to ask to which accuracy one expects the 
scaling law for elliptic flow to hold. In particular, are the scaling 
factors of 2 and 3 indeed excluding any higher Fock states in the hadrons?
A recent study found that higher Fock states in an expansion
 $ |p\rangle = a_0 |uud \rangle + a_1 | uudg \rangle + 
  \ldots $
could actually be accommodated \cite{MFB:05}. This is in fact trivial for 
the single-hadron yields from a thermal parton spectrum. They do not
change if additional partons are allowed to coalesce.
The situation changes for elliptic flow. Higher Fock states with $n$
partons come with their own scaling factor $n$ which seems to destroy 
the scaling with the number of valence quarks. However a numerical 
evaluation shows that the correction is surprisingly small.
Fig.\ \ref{fig:hf} shows the expected violation of the scaling law using 
the new asymmetry variable $A=(B-M)/(B+M)$ where $B$ and $M$ are the 
{\it scaled} elliptic flow of a meson and a baryon respectively \cite{MFB:05}.
Generally the violations are smaller than
5\%. New data from STAR analyzes scaling violations in the data and
finds them to have the predicted sign and order of magnitude 
\cite{Sorensen:05}. After this study it is 
clear that there is some room to accommodate gluons or sea quarks 
during the recombination process.
Further investigations in this direction are necessary.

\section{Hadron Correlations}

One crucial simplification implemented so far is a factorization of any 
$n$-parton Wigner function into a product of independent single parton 
distributions
 $ W_{1,\ldots,n} = \prod_{i=1}^n w_i$.
By definition this factorization does not permit any correlations between 
partons. Consequently, no hadron correlations can emerge via recombination.
Originally this factorization was chosen for simplicity and it was 
justified because single inclusive hadron spectra could be described 
very well. It has been shown in \cite{FMB:04} that modifications of 
this factorization including correlations between partons lead to 
correlations between hadrons after recombination. The quality of the 
single hadron spectra does not suffer.
The source of jet-like correlations in the QGP is the strong
coupling of jets to the medium.
The energy loss is estimated to be up to 14 GeV/fm for a 10 GeV parton 
\cite{Wang:04jq}. Quenched jets lead to a considerable local heating and 
create a hot spot inside the fireball. Moreover, the directional information 
of the jet is preserved. Partons of such a hot spot exhibit jet-like 
correlations. 

The next oder extension of the correlation-free factorization in \cite{FMB:04}
assumes that correlations still are a small effect and that one can 
restrict them to 2-particle correlations $C_{ij}$ (see also \cite{FB:05}). 
Then a 4-parton Wigner function can be written
\begin{equation}
  W_{1234} \approx w_1 w_2 w_3 w_4 \big( 1 + \sum_{i<j}C_{ij} \big).
  \label{eq:newfact}	
\end{equation}
The correlation functions $C_{ij}$ between parton $i$ and parton $j$ are 
non-vanishing only in a subvolume $V_c$ of the fireball.
A Gaussian ansatz $C_{ij} \sim c_0 e^{-(\phi_i-\phi_j)^2/(2 \phi_0^2)}$
is chosen to describe correlations in azimuthal angle.
The 2-meson yield is given by a convolution of the partonic Wigner function
$W_{1234}$ with the Wigner functions $\Phi_A$, $\Phi_B$ of the mesons with
an additional integration over the hadronization hypersurface \cite{FMB:04}. 
It is assumed that the correlation strength $c_0 \ll 1$ which 
permits omitting quadratic terms like $c_0^2$ or $c_0 v_2$.

We can now study the associated yield $Y_{AB}$ for a trigger hadron
$A$ in a given kinematic window as a function of the relative azimuthal
angle $\Delta \phi$ between the two. One finds
\begin{equation}
  2\pi N_A Y_{AB} (\Delta\Phi) = Q \hat c_0 e^{-(\Delta\Phi)^2/(2\phi_0)^2}
  N_A N_B.
  \label{eq:res}
\end{equation}
The $N_i$ are single particle yields in the kinematic window of the trigger 
meson or associated meson and $\hat c_0 = c_0 V_c/(\tau A_T)$ where
$\tau A_T$ is the hadronization volume.
The factor $Q=4$ (for two mesons) indicates an enhancement of the correlations 
in the hadron phase compared to the parton phase. The effect is similar to
the amplification of elliptic flow by the number $n$ of valence quarks. 
$Q$ counts the number of possible correlated pairs between the $n_A$ 
(anti)quarks of meson $A$ and the $n_B$ (anti)quarks of meson $B$. 
In the weak correlation limit only single correlations are counted.
Apparently one has
 $ Q=n_A n_B,$
thus $Q=6$ for a meson-baryon pair and $Q=9$ for a baryon-baryon pair.

\begin{figure}
\begin{center}
\epsfig{file=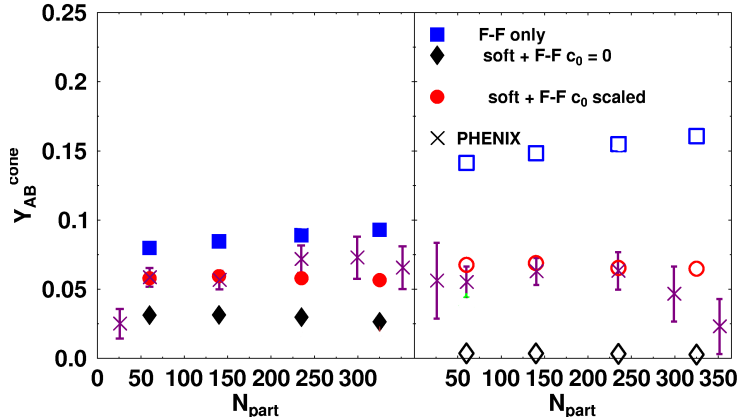,width=10cm} \\
\caption{\label{fig:res} The associated yield $Y_{AB}^{\mathrm cone}$ for
baryon triggers (right panel) and meson triggers (left panel) as a function
of $N_{\mathrm part}$. Squares: fragmentation only; diamonds: fragmentation
and recombination with $\hat c_0=0$;
circles: the same with $\hat c_0=0.08\times 100/N_{\mathrm part}$;
data: PHENIX \cite{phenix:04corr}.}
\end{center}
\end{figure}

Fig.\ \ref{fig:res} shows the associated yield of hadrons integrated 
in azimuthal angle around the near side ($\Delta\phi =0$)
for the case that the trigger is a baryon (proton or antiproton) and 
a meson (pion or kaon) for different centralities. 
The kinematic window is 2.5 GeV/$c \le P_{TA}\le$4.0 GeV/$c$ for trigger 
particles and 1.7 GeV/$c\le P_{TB}\le$2.5 GeV/$c$ for associated particles, 
and $|y_A|$, $|y_B|<0.35$.
Fig.\ \ref{fig:res} shows the associated yield with only fragmentation,
and fragmentation and recombination both taken into account together
with PHENIX data \cite{phenix:04corr}. A good description of the data can
be reached assuming a constant correlation volume.
The parameters used for the fireball are the same that lead to 
a good description of single hadron spectra and elliptic flow
\cite{FMNB:03prc}.

\ack I would like to thank my collaborators B.\ M\"uller, S.\ A.\ Bass and 
C.\ Nonaka. This work was supported by DOE grant DE-FG02-87ER40328. RIKEN,
Brookhaven National Laboratory and the U.S. Department of Energy,
grant DE-AC02-98CH10886, provided facilities for the completion of 
this work. Many thanks to the organizers of Hard Probes 2006 for inviting
me to this conference.

\end{document}